\documentclass{IEEEtran}

\usepackage[T1]{fontenc}
\usepackage{graphicx}% http://ctan.org/pkg/graphicx
\graphicspath{{figures/}}
\usepackage{epstopdf}
\usepackage{amssymb,amsmath,amsfonts}
\usepackage{url,changebar,bm,xspace,dsfont}
\usepackage{theorem}
\usepackage{epsfig}
\usepackage{epstopdf}
\usepackage{cite}
\usepackage{verbatim}
\usepackage{bbm} 																											%
\usepackage[usenames,dvipsnames]{color}
\usepackage{flushend}
\usepackage{tikz}
\usepackage{stackrel}
\usepackage{footnote}
\usepackage{dblfloatfix}
\usepackage[normalem]{ulem} % to have strikethrough  words
\usepackage{lipsum}% http://ctan.org/pkg/lipsum
\usepackage{multirow}																									%
\usepackage[section,verbose]{placeins} 																		%
\usepackage{tikz}																											%
\usetikzlibrary{arrows,automata,matrix,fit,positioning,calc}             								%
\usepackage{pgfplots}																									%
\usepackage[latin1]{inputenc}																						%
\usepackage{verbatim}																									%
\usepackage{moreverb}
\usepackage{placeins}
\usepackage[linesnumbered,lined,commentsnumbered]{algorithm2e}
\usepackage{algorithmicx}
\algdef{SE}[DOWHILE]{Do}{doWhile}{\algorithmicdo}[1]{\algorithmicwhile\ #1}%
\algnewcommand{\algorithmicgoto}{\textbf{Go to}}%
\algnewcommand{\Goto}[1]{\algorithmicgoto~\ref{#1}}%

\newcommand{\BD}[1]{\textcolor{blue}{{#1}}}

\newcommand{\EG}[1]{\textcolor{blue}{{#1}}}

\def\x{\bm{x}}\def\y{\bm{y}}\def\w{\bm{w}}\def\v{\bm{v}}\def\0{\bm{0}}\def\bu{\bm{u}}\def\h{\bm{h}}

\newtheorem{remark}{Remark}
\newtheorem{definition}{\bfseries Definition}
\newtheorem{theorem}{\bfseries Theorem}

\newtheorem{corollary}{\bfseries Corollary}
\newtheorem{example}{\bfseries Example}

\def\I{\bm{I}}
\def\I{\bm{\mathrm{I}}}

\def\w{\bm{w}}\def\x{\bm{x}}\def\y{\bm{y}}\def\v{\bm{v}}\def\z{\bm{z}}

\usepackage{setspace}

\title{\LARGE Optimal control of linear systems with limited control actions: threshold-based event-triggered control
}

\author{\IEEEauthorblockN{Burak Demirel, Euhanna Ghadimi, Daniel E. Quevedo and Mikael Johansson}
\thanks{
			B. Demirel and D. E. Quevedo are with the Faculty of Electrical Engineering and Information Technology, The University of Paderborn, Warburger Str. 100, 33098 Paderborn, Germany (e-mail: burak.demirel@upb.de; dquevedo@ieee.org). }
			\thanks{E. Ghadimi is with Huawei Technologies Sweden AB, Skalhogatan 9-11 box 54, SE-164 94, Kista, Sweden (e-mail: euhanna.ghadimi@huawei.com).}
			\thanks{M. Johansson is with ACCESS Linnaeus Center, School of Electrical Engineering, KTH Royal Institute of Technology, Osquldas v\"{a}g~10, SE 10044 Stockholm, Sweden (e-mail: mikaelj@ee.kth.se). } 
			}

\begin{document}

\maketitle

\begin{abstract}
We consider a finite-horizon linear-quadratic optimal control problem where only a limited number of control messages are allowed for sending from the controller to the actuator. To restrict the number of control actions computed and transmitted by the controller, we employ a threshold-based event-triggering mechanism that decides whether or not a control message needs to be calculated and delivered. Due to the nature of threshold-based event-triggering algorithms, finding the optimal control sequence requires minimizing a quadratic cost function over a non-convex domain. In this paper, we firstly provide an exact solution to the non-convex problem mentioned above by solving an exponential number of quadratic programs. To reduce computational complexity, we, then, propose two efficient heuristic algorithms based on greedy search and the Alternating Direction Method of Multipliers (ADMM) method. Later, we consider a receding horizon control strategy for linear systems controlled by event-triggered controllers, and we also provide a complete stability analysis of receding horizon control that uses finite horizon optimization in the proposed class. Numerical examples testify to the viability of the presented design technique. 
\end{abstract}

% ====================================================================================================
%
% Keywords
%
% ====================================================================================================

%====================================================================================================
%
% INTRODUCTION
%
%====================================================================================================

\section{Introduction}\label{sec:Introduction}

The problem of determining optimal control policies for discrete-time linear systems has been extensively investigated in the literature; see, e.g.,~\cite{AnM:90, Ber:00, Ast:06}. The standard optimal control problem assumes that an unlimited number of control actions is available at the actuator. Although this assumption is valid for many applications, it does not hold for some specific scenarios where communication and computation resources become scarce; for example, (a) control systems with constrained actuation resources~\cite{GaM:12, HGM:13, CGH+:13}, (b) control systems with shared processor resources~\cite{Gup:10}, or (c) information exchange over a shared communication channel~\cite{DGQ+:16}.

To reduce the communication burden between the controller and the actuator, Imer and Ba\c{s}ar~\cite{ImB:06} introduced a constraint on the number of control actions while designing optimal control policies for linear scalar systems. Later, Bommannavar and Ba\c{s}ar~\cite{BoB:08} and Shi~et al.~\cite{SYC:13} extended the work of~\cite{ImB:06} to a class of higher-order systems. The problem, proposed in~\cite{ImB:06}, can be also formulated as a cardinality-constrained linear-quadratic control problem. The introduction of cardinality constraint changes the standard linear-quadratic control problem from a quadratic programming (QP) to a mixed-integer quadratic programming (MIQP) problem. As it has been proved in~\cite{Gao:05}, it is an NP-complete problem. Therefore, Gao and Lie~\cite{GaL:11} provided an efficient branch and bound algorithm to compute the optimal control sequence. Since the cardinality constraint is highly non-convex, a convex relaxation of this problem, based on $\ell_{1}$-regularized $\ell_{2}$ optimization, was considered in the literature; see, e.g.,~\cite{NaQ:11, GaM:13, NQO:14, ADD+:14}. Although $\ell_{1}$-norm regularization of the cost function is an effective way of promoting sparse solutions, no performance guarantees can be provided in terms of the original cost function due to the transformed cost function.

Event- and self-triggered control systems have been broadly used in the literature to reduce the amount of communication between the controller and the actuator while guaranteeing an attainable control performance; see, e.g.,~\cite{AsB:02, AnH:14, GAD+:14, GoH:15, DGQ+:15}. As distinct from \textit{sparse control} techniques as mentioned earlier, the event- and self-triggered control require the design of both a feedback controller that computes control actions and a triggering mechanism that determines when the control input has to be updated. A vast majority of the works in the literature first designed a controller without considering sparsity constraint, and then, in the subsequent design phase, they developed the triggering mechanism for a fixed controller. In contrast, another line of research concentrates on the design of feedback control law while respecting a predefined triggering condition. 

\BD{
The design of event- and self-triggered algorithms can be extremely beneficial in the context of model predictive control (MPC) strategies; see, e.g.,~\cite{SLH:10,EDK:10,EDK:11,LHJ:13,EQP+:15,HeS:15,BHA:16,VKF+:09,LiS:14}. MPC is a control scheme that solves a finite-horizon optimal control problem at each sampling instant and only applies the first element of the resulting optimal control input trajectories. The use of event-triggered algorithms, therefore, reduces the frequency of solving optimization problems and transmitting control actions from the controller to the actuator, and, consequently, saves computational and communication resources. Lehmann et al.~\cite{LHJ:13} proposed an event-based strategy for discrete-time systems under additive, bounded disturbances. The controller only computes a new control command whenever the difference between actual state and predictive state exceeds a threshold. Sijs et al.~\cite{SLH:10} combined state estimation with MPC in order to design an event-based estimation framework. The authors of~\cite{EDK:10,EDK:11} combined MPC with event-triggered sampling strategies based on the ISS concept. As different from the aforecited works, the authors of~\cite{AnH:14,GAD+:14,GoH:15} studied infinite horizon quadratic cost. All works discussed above focus on discrete-time linear/non-linear systems; however, there are also a substantial number of works, which considers continuous-time systems, in the literature; see, e.g.,~\cite{VKF+:09,VaF:11,LiS:14}. For instance, the authors of~\cite{VKF+:09,VaF:11} investigated the stability of event-based MPC algorithms for continuous-time nonlinear systems yet they did not consider disturbance. Later, Li and Shi~\cite{LiS:14} studied the MPC problem for continuous-time nonlinear systems subject to bounded disturbances. 
}

\subsubsection*{Contributions}
In this paper, we formulate a finite-horizon optimal event-triggered control problem where a threshold-based event-triggering algorithm dictates the communication between the controller and the actuator. Then, we propose various algorithms to compute a control action sequence that provides optimal and sub-optimal solutions for this problem, which is, in general, hard to solve due to two main reasons: (a) it has the combinatorial nature since the decisions are binary variable (i.e., transmit or not transmit), and (b) introduction of a threshold-based triggering condition leads to optimizing a convex cost function over a non-convex domain. The main contributions of this paper are threefold:
\begin{itemize}
	\item[(i)] we show that the optimal solution of the control problem, mentioned above, can be determined via solving a set of quadratic programming problems;
	\item[(ii)] we provide an efficient heuristic algorithm based on ADMM to design a control input sequence that provides a sub-optimal solution for the finite-horizon optimal event-triggered control problem;
	\item[(iii)] we describe the receding horizon implementation of the event-triggered control algorithm, including a proof of practical stability.
\end{itemize}

\subsubsection*{Outline}
The remainder of this paper is organized as follows. Section~\ref{sec:FHOETC} formulates the event-triggered finite-horizon LQ control problem and introduces assumptions. Section~\ref{sec:opt_control} provides a simple procedure for designing optimal control laws to minimize the linear-quadratic cost function while Section~\ref{sec:subopt_control} presents two heuristic methods that usually achieves tolerable sub-optimal performance while significantly reducing the computational complexity. In Section V, a receding horizon control scheme with event-triggered algorithm is presented. Section VI demonstrates the effectiveness and advantages of the presented approach while Section VII concludes the paper.

\subsubsection*{Notation}
We write $\mathbb{N}$ for the positive integers, $\mathbb{N}_{0}$ for $\mathbb{N}\cup\{0\}$, and $\mathbb{R}$ for the real numbers. Let $\mathbb{R}_{\succeq 0}^{n}$ denote the set of non-negative real vectors of dimension $n$, and $\mathbb{R}^{n}$ be the set of real vectors of dimension $n$. Vectors are written in bold lower case letters (e.g., $\bm{u}$ and $\bm{v}$) and matrices in capital letters (e.g., $A$ and $B$). If $\bm{u}$ and $\bm{v}$ are two vectors in $\mathbb{R}^{n}$, the notation $\bm{u}\leq\bm{v}$ corresponds to component-wise inequality. The set of all real symmetric positive semi-definite matrices of dimension $n$ is denoted by $\mathbb{S}_{\succeq 0}^{n}$. We let $\mathbf{0}_{n}$ be the $n$--dimensional column vectors of all zeros, $\mathbf{1}_{n}$ be the vectors of all ones. The Kronecker product of two matrices (e.g., $A$ and $B$) is denoted by $A \otimes B$. For any given $\bm{x}\in\mathbb{R}^{n}$, the $\ell_{\infty}$--norm is defined by $\parallel\bm{x}\parallel_{\infty}=\max\limits_{1\leq i\leq n}\vert x_{i} \vert$. For a square matrix $A$, $\lambda_{\max}(A)$ denotes its maximum eigenvalue in terms of magnitude. The notation$\{x_{k}\}_{k\in\mathcal{K}}$ stands for $\{x(k) : k\in\mathcal{K}\}$, where $\mathcal{K}\subseteq\mathbb{N}_{0}$. The power set of any set $\mathcal{N}$, written $\mathsf{P}(\mathcal{N})$, is the set of all subsets of $\mathcal{N}$, including the empty set and $\mathcal{N}$ itself. The cardinality of a set denoted by $\vert\mathcal{N}\vert$.

\section{Finite Horizon Optimal Event-Triggered Control}\label{sec:FHOETC}

\begin{figure}[!t]\centering
	\includegraphics[scale=0.30]{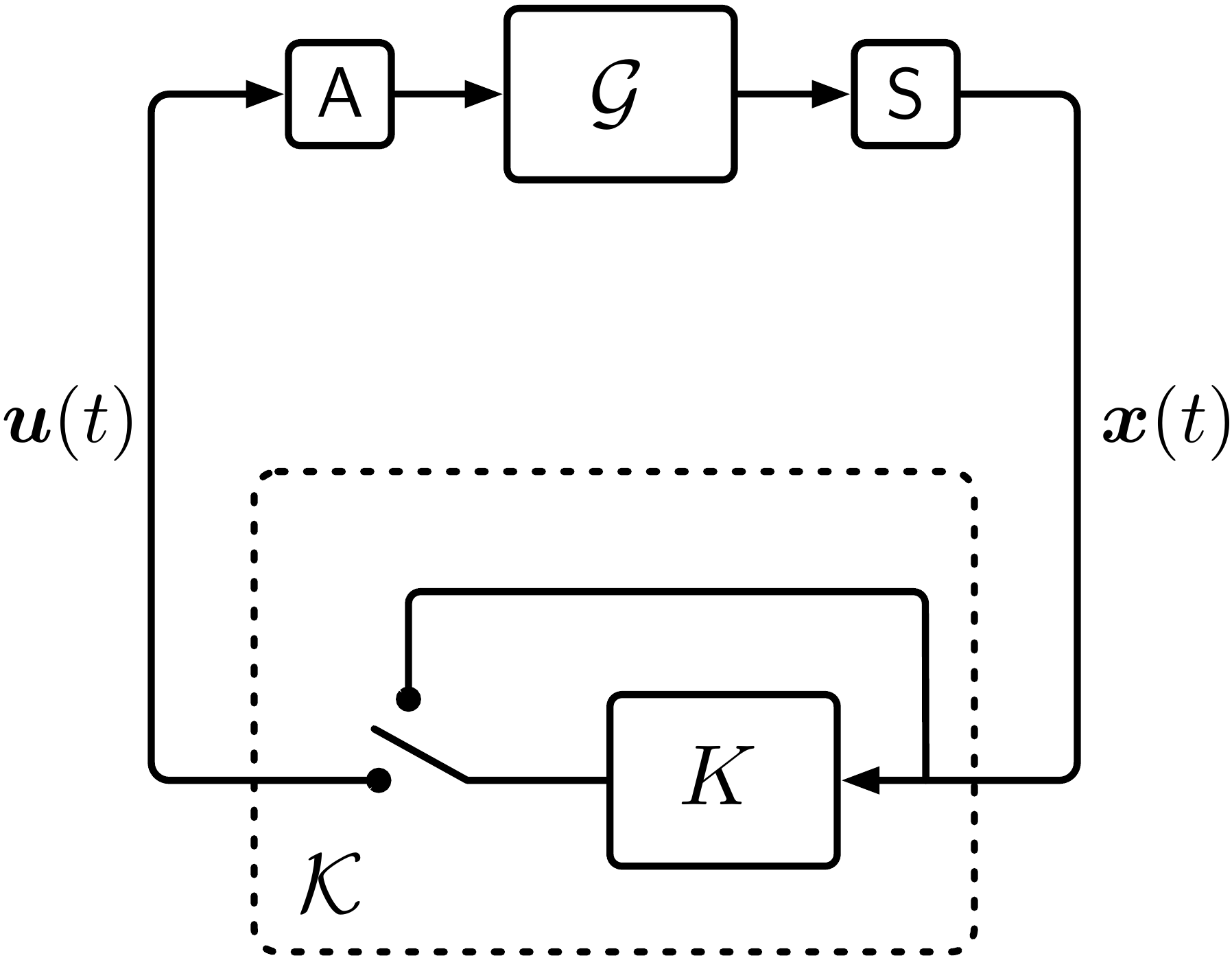}
	\caption{Event-triggered control system with the process $\mathcal{G}$, the actuator $\mathsf{A}$, the sensor $\mathsf{S}$ and the controller $\mathcal{K}$.} \label{fig:Block_diagram}
\end{figure}

We consider the feedback control loop, depicted in Fig.~\ref{fig:Block_diagram}. The dynamics of the physical plant $\mathcal{G}$ can be described by the discrete-time linear time-invariant system:
\begin{align}
	\mathcal{G} : \quad \x(t+1) = A\x(t) + B\bu(t)\;, \quad \x(0) = \x_{0}\;, 	\label{eqn:Discrete_time_system}
\end{align}
where $\x(t)\in\mathbb{R}^{n}$ is the state variable at time instant $t$, $\bu(t)\in\mathbb{R}^{m}$ is the control input at time instant $t$, $A\in\mathbb{R}_{}^{n\times n}$ and $B\in\mathbb{R}_{}^{n\times m}$ are system matrices of appropriate dimensions, and $\x_{0}\in\mathbb{R}^{n}$ is the given initial condition. The pair $(A,B)$ is assumed to be stabilizable, and $A$ is not necessarily Schur stable. 

In this paper, we assume that the sensor $\mathsf{S}$ takes periodic noise-free samples of the plant state $\x(t)$ and transmits these samples to the controller node. The controller $K$ is event-triggered, and it computes new control commands and transmits them to the actuator $\mathsf{A}$ only at times when $\x(t)\in \mathcal{S} \triangleq \mathbb{R}_{}^{n}\setminus\mathcal{C}_{0}^{}$ with  %t\in\mathbb{T}_{N}^{}\triangleq\{t\in\mathbb{N}_{0}^{} : t<N\}$
\begin{align}
	\mathcal{C}_{0}^{}\triangleq\{\x\in\mathbb{R}_{}^{n} : \parallel \x \parallel_{\infty}^{} < \varepsilon \} \;, \label{eqn:event_triggered_set}
\end{align}
for a given threshold $\varepsilon > 0$.
In case $\x(t)\in\mathcal{C}_{0}^{}$, the controller does not compute any new control actions, and the actuator input to the plant is set to zero:
\begin{align}
	\bu(t) = \mathbf{0}_{m}^{}, \quad \forall t\in\mathcal{T}, \label{eqn:event_triggered_constraint}
\end{align}
where 
\begin{align}\label{eqn:event_triggered_constraint_time}
	\mathcal{T}\triangleq\{t\in\mathbb{N}_{0}^{}; t < N : \x(t)\in\mathcal{C}_{0}^{}\} \;.
\end{align}
It is worth noting that $\mathcal{T}$ denotes a set of time instants at which no control computations are needed and, in turn, the plant runs open loop. Notice that $\mathcal{T}$ is not a given set of variables; in contrast, it is generated by the initial state $\x_{0}^{}$ and the tentative control actions $\bu(t)$ for all $t\in\{0,\cdots,\mathrm{T}-1\}$. Hence, to determine the set $\mathcal{T}$, it is necessary to compute the tentative control actions $\bu(t)$. In other words, the set of time instants $\mathcal{T}$ is strongly coupled to the tentative control inputs $\bu(t)$.

Throughout this paper, we aim at designing an \emph{admissible} optimal control sequence $\boldsymbol{\pi} = \{\bu(0), \bu(1), \cdots, \bu(N-1)\}$ to minimize the quadratic cost function:
\begin{multline}
	J(\x(0),\boldsymbol{\pi}) = \x^{\intercal}(N)P\x(N) \\ + \sum_{t=0}^{N-1} \big(\x^{\intercal}(t)Q\x(t) + \bu^{\intercal}(t)R\bu(t) \big) \;, \label{eqn:cost_function}
\end{multline}
where the matrices $Q$ and $P$ are symmetric and positive semi-definite while $R$ is symmetric and positive definite. Then, we consider the constrained finite-time optimal control problem:
\begin{align}	
	\begin{array}{rll}
		J_{}^{\star}(\x(0)): & \stackrel[\boldsymbol{\pi}]{}{\text{minimize}}	 & J(\x(0),\boldsymbol{\pi}) \\
										 & \text{subject to} & \x(t+1) = A\x(t) + B\bu(t),  \\
										 & & \hspace{11mm} \forall t\in\{0,\cdots,N-1\}, \\
										 & & \x(0) = \x_{0}^{}, \\
									     & & \bu(t) = \mathbf{0}_{m}^{}, \quad \forall t\in\mathcal{T}.
	\end{array}	 
	\label{eqn:opt_problem}
\end{align}
Recall that the set of time instants, at which no computations (and also transmissions) are needed, i.e., $\mathcal{T}$, is defined in~\eqref{eqn:event_triggered_constraint_time}. 

\begin{remark}
Note that the event-triggered control problem, described in Section~\ref{sec:FHOETC}, is a particular class of hybrid systems. One may, therefore, convert it into an equivalent mixed logical dynamical (MLD) system that represents the system by using a blend of linear and binary constraints on the original variables and some auxiliary variables; see, e.g.,~\cite{BTM:00,HSB:01}. In this paper, instead of using the \EG{MLD} system formulation, we will exploit geometric properties of the problem at hand, which allows us to propose an optimization problem for optimal finite-horizon control and establish stability result for its receding horizon implementation.
\end{remark}

\subsubsection*{Discussion}
Although the problem~\eqref{eqn:opt_problem} resembles the cardinality-constrained optimal control problem \BD{(proposed in~\cite{ImB:06, BoB:08, SYC:13, Gao:05, GaL:11, NaQ:11, GaM:13, NQO:14, ADD+:14}),} there is a fundamental difference between them, which stems from the question whether scheduling is an exogenous input or autonomously generated? While solving the cardinality-constrained control problem, one needs to optimize both control action and scheduling sequence, which are strongly coupled to each other in the optimization process. On the other hand, the event-triggered control systems are switched systems with internally forced switchings. In these problems, scheduling sequences are generated implicitly based on the evolution of the state $\x(t)$ and the control signal $\bu(t)$. The major difficulty of this problem is that scheduling depends on the particular initial condition $\x_{0}^{}$ and the tentative control input $\bu(t)$, and cannot be explicitly determined unless a specific control signal was given; see the survey paper~\cite{ZhA:14} and references therein.

\section{Computation of optimal control actions}\label{sec:opt_control}

In this section, we concentrate on finding the control input sequence $\boldsymbol{\pi}$ that solves the finite-horizon optimal control problem, proposed in~\eqref{eqn:opt_problem}. Therefore, we present a framework which is based on dividing the non-convex domain into convex sub-domains. This is a simple yet effective procedure to follow; however, the number of convex optimization problems, which needs to be solved, grows exponentially in the length of the control horizon.

\begin{figure}\centering
	\includegraphics[scale=0.27]{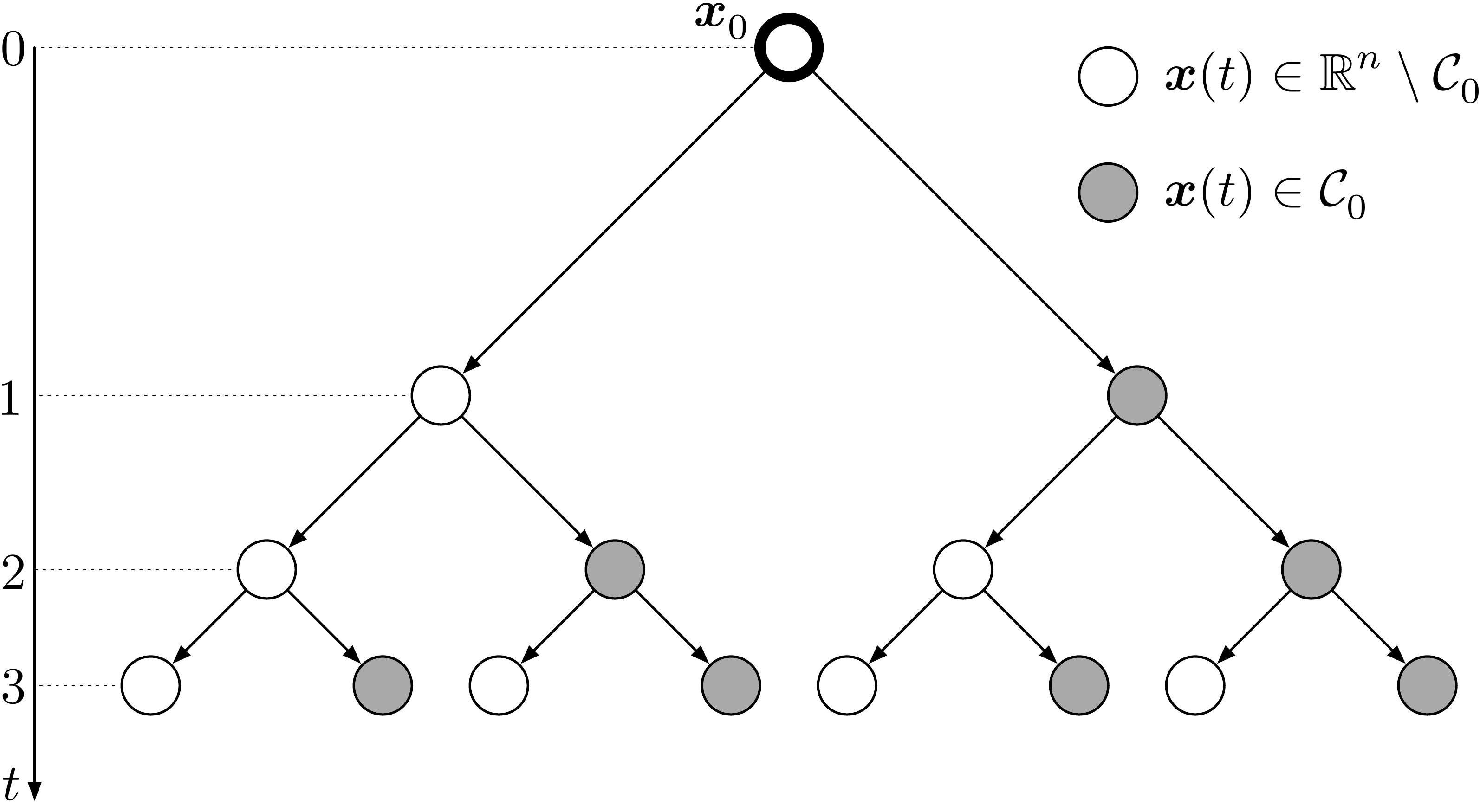}
	\caption{A decision tree diagram. Denote $\mathcal{T}\triangleq\{t\in\mathbb{N}_{0}^{}; t<N : \x(t)\in\mathcal{C}_{0}^{}\}$. This means that $\mathcal{T}$ represents a set of time instances at which a new control signal does not need to be computed and transmitted from the controller to the actuator. In the figure, dark colored circles represent the case $\x(t)\in\mathcal{C}_{0}^{}$ for any $t\in\mathcal{T}$ whereas light colored circles represent the case $\x(t)\in\mathbb{R}^{n}\setminus\mathcal{C}_{0}^{}$ for any $t\notin\mathcal{T}$. For instance, when $N=4$, the set $\mathcal{T}$ gets values in $\mathsf{P}(\{1,2,3\})=\big\{\emptyset,\{1\},\{2\},\{3\},\{1,2\},\{1,3\},\{2,3\},\{1,2,3\}\big\}$.} \label{fig:Decision_tree}
\end{figure}

The optimal control problem~\eqref{eqn:opt_problem} is hard to solve due to the restriction on the control space. Nevertheless, without loss of generality, it is possible to convert this problem to a set of optimization problems of the form
\begin{align}	
	\begin{array}{rll}
		J_{\mathcal{T}}^{\star}(\x(0)): & \stackrel[\boldsymbol{\pi}]{}{\text{minimize}}	 & J(\x(0),\boldsymbol{\pi}) \\
										 & \text{subject to} & \x(t+1) = A\x(t) + B\bu(t),  \\
										 & & \hspace{11mm} \forall t\in\{0,1,\cdots,N-1\}, \\
									     & & \x(t)\in\mathcal{C}_{0}^{}, ~ \forall t\in\mathcal{T}, \\
										 & & \x(t)\in\mathbb{R}_{}^{n}\setminus\mathcal{C}_{0}^{}, ~ \forall t\notin\mathcal{T}, \\
										 & & \bu(t) = \0_{m}^{}, ~ \forall t\in\mathcal{T}, \\
										 & & \x(0) = \x_{0},
	\end{array}	 
	\label{eqn:opt_problem_noncvx}
\end{align}
where $\mathcal{T}\in\mathsf{P}(\mathcal{N})$ with $\mathcal{N}=\{1,2,\cdots,N-1\}$. To obtain the optimal solution of the problem~\eqref{eqn:opt_problem}, it is necessary to solve the problem~\eqref{eqn:opt_problem_noncvx} for all subsets of the power set of $\mathcal{N}$, i.e., $\mathsf{P}(\mathcal{N})$, and select the one providing the lowest cost; see~Fig.~\ref{fig:Decision_tree}. More precisely, this can be written as:
\begin{align}	
	\begin{array}{rll}
		J_{}^{\star}(\x(0)): & \stackrel[\mathcal{T}\in\mathsf{P}(\mathcal{N})]{}{\text{minimize}}	 & J_{\mathcal{T}}^{\star}(\x(0)) \;.
	\end{array}	 
%	\label{eqn:opt_problem_noncvx}
\end{align}

The number of sub-problems, which are needed to be solved, grows exponentially with the control horizon $N$ (i.e., $|\mathsf{P}(\mathcal{N})| = 2^{N-1}$). Bear in mind that since these sub-problems are independent of each other, they can be also solved in parallel. In addition to its combinatorial nature, the problem~\eqref{eqn:opt_problem_noncvx} requires the optimization of a convex function over a non-convex domain. In the literature, there exist some available tools to solve the convex objective non-convex optimization; see, e.g.,~\cite{Mic:13, BiM:14, DTB:16}. 

\begin{figure}[!t]\centering
	\includegraphics[scale=0.9]{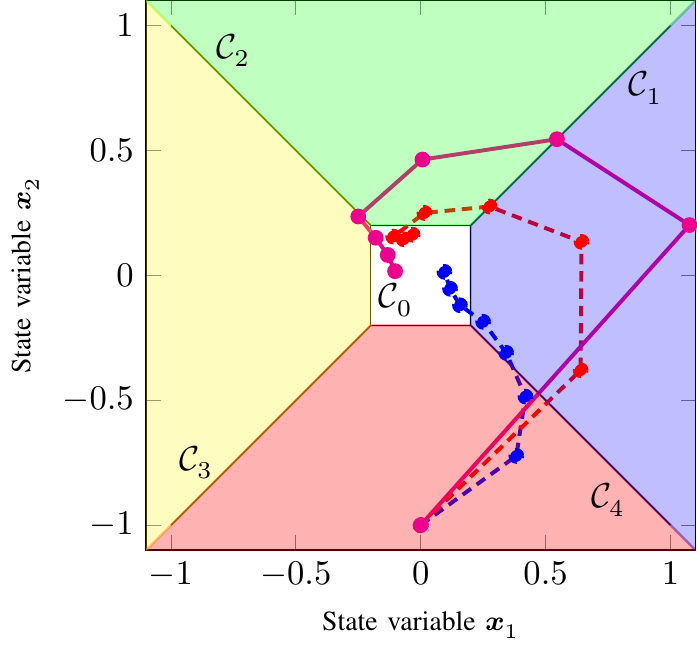}
  	\caption{The non-convex set $\mathcal{S}\triangleq\mathbb{R}^{2}\setminus\mathcal{C}_{0}^{}$ can be divided into four convex regions, which are illustrated by different colors. For a given scheduling sequence, i.e., $\mathcal{T} = \{6, 7\}$, the optimal control inputs can be computed by solving the optimization problem~\eqref{eqn:opt_problem_cvx} for all possible combinations of five convex sets. For $N=7$, finding optimal control actions requires solving $15,625$ problems, and selecting the minimum control cost among $2,650$ feasible solutions. In Fig.~\ref{fig:Example_1}, the blue curve represents the state trajectory for $\Sigma_{1}^{}=\{ 4, 4, 4, 1, 1, 0, 0 \}$, the red curve represents the state trajectory for $\Sigma_{2}^{}=\{ 4, 1, 1, 2, 2, 0, 0 \}$, and the magenta curve denotes the state trajectory for $\Sigma_{3}^{}=\{ 4, 1, 2, 2, 3, 0, 0 \}$.}
  	\label{fig:Example_1} 
\end{figure}

As the feasible region $\mathcal{S}$ can be expressed as a finite union of polyhedra, the disjunctive formulation can be applied. In particular,  assuming $n$-dimensional problem instance~\eqref{eqn:opt_problem}, its feasibility region can be divided into $2n+1$ disjunctive polyhedra\footnote{See Fig.~\ref{fig:Example_1} for an example of two dimensional problem where each disjunctive set is shown in different colors.}. Following the construction above, we write:
\begin{align}
	\mathcal{S} \triangleq \bigcup_{p = 1}^{2n} \mathcal{C}_{p}^{} \;, \label{eqn:Union_set}
\end{align}
where $n$ is the state dimension, and $\mathcal{C}_{p}^{}\subset\mathbb{R}_{}^{n}$ are full-dimensional polyhedra, i.e.,
\begin{align*}
	\mathcal{C}_{p}^{} \triangleq \big\{ \x\in\mathbb{R}_{}^{n} : T_{p}^{}\x \leq \boldsymbol{d}_{}^{} \big\} \;,
\end{align*}
for some $T_{p}^{}\in\mathbb{R}_{}^{(2n-1)\times n}$ and $\boldsymbol{d}_{}^{}\in\mathbb{R}_{}^{2n-1}$. 

We define a piece-wise constant function of time $\sigma(t)$, which takes on values in $\{0,1,\cdots,2n\}$ and whose value $p$ determines, at each time $t\in\{0,\cdots,\mathrm{T}-1\}$, the state variable $\x(t)$ to belong to the interior of the polyhedron $\mathcal{C}_{p}^{}$. We divide the non-convex set $\mathcal{S}$ into convex subsets $\mathcal{C}_{p}$ with $p\in\{0,1,\cdots,2n\}$ and, at each time step $t\in\{0,\cdots,\mathrm{T}-1\}$, we choose only one active set. The switching signal,  
\begin{align}
	\sigma(t) = 
	\begin{cases}
	 0 & \text{if} ~ t\in\mathcal{T}, \\
	 p & \text{otherwise},
	\end{cases}
\end{align}
gives a sequence $\Sigma=\{\sigma(0),\cdots,\sigma(N-1)\}$. Then, for given $\mathcal{T}$ and $\Sigma$, we rewrite the optimization problem~\eqref{eqn:opt_problem_noncvx} as
\begin{align}	
	\begin{array}{rll}
		J_{\Sigma}^{\star}(\x(0)): & \stackrel[\boldsymbol{\pi}]{}{\text{minimize}}	 & J(\x(0),\boldsymbol{\pi}) \\
										 & \text{subject to} & \x(t+1) = A\x(t) + B\bu(t),  \\
										 & & \hspace{6mm} \forall t\in\{0,1,\cdots,N-1\}, \\
										 & & \x(t)\in\mathcal{C}_{\sigma(t)}, ~ \forall \sigma(t)\in\Sigma, \\
										 & & \bu(t) = \0_{m}^{}, ~ \forall t\in\mathcal{T}, \\
										 & & \x(0) = \x_{0}.
	\end{array}	 
	\label{eqn:opt_problem_cvx}
\end{align}

It is necessary to solve all possible combination of problem~\eqref{eqn:opt_problem_cvx} to determine the global optimal solution. Particularly, the problem~\eqref{eqn:opt_problem} can be solved as a sequence of $(2n+1)^{N-1}$ quadratic programming (QP) problems. Hence, we have the following series of optimization problems:
  \begin{align}\label{eqn:opt_sequence_problem}	
	\begin{array}{rll}
		J_{}^{\star}(\x(0)): & \stackrel[\Sigma\in\{0,\cdots,2n\}_{}^{N-1}]{}{\text{minimize}}	 & J_{\Sigma}^{\star}(\x(0)) \;.
	\end{array}	 
\end{align}

It is worth noting that even though the number of QP sub-problems grows exponentially with $N$ and $n$, all these problems are independent of each other; therefore, one can parallelize the problems and  reduce the computation time. 

\begin{corollary}
	The global optimal solution of the control problem~\eqref{eqn:opt_problem} can be computed by solving a set of convex quadratic programming problems of the form~\eqref{eqn:opt_problem_cvx}, in parallel, and selecting the solution that provides the lowest control cost among all feasible solutions.
\end{corollary}

The following examples should give the reader a better understanding of the problem.

\begin{example}
Let $n=2$ and $\mathcal{S}=\mathbb{R}^{2}\setminus\mathcal{C}_{0}^{}$ with $\mathcal{C}_{0}^{} = \{ (x_{1}^{},x_{2}^{})\in\mathbb{R}^{2} : -\varepsilon < x_{1}^{} < \varepsilon, -\varepsilon < x_{2}^{} < \varepsilon \}$. The non-convex set $\mathcal{S}$ can be divided into four convex subsets:
\begin{align*}
	\mathcal{C}_{1}^{} &= \{ (x_{1}^{},x_{2}^{})\in\mathbb{R}^{2} : x_{1}^{} \geq x_{2}^{}, x_{1}^{} \geq -x_{2}^{}, x_{1}^{} \geq \varepsilon \} \;, \\
	\mathcal{C}_{2}^{} &= \{ (x_{1}^{},x_{2}^{})\in\mathbb{R}^{2} : x_{1}^{} \leq x_{2}^{}, x_{1}^{} \geq -x_{2}^{}, x_{2}^{} \geq \varepsilon \} \;, \\
	\mathcal{C}_{3}^{} &= \{ (x_{1}^{},x_{2}^{})\in\mathbb{R}^{2} : x_{1}^{} \leq x_{2}^{}, x_{1}^{} \leq -x_{2}^{}, x_{1}^{} \leq -\varepsilon \} \;, \\
	\mathcal{C}_{4}^{} &= \{ (x_{1}^{},x_{2}^{})\in\mathbb{R}^{2} : x_{1}^{} \geq x_{2}^{}, x_{1}^{} \leq -x_{2}^{}, x_{2}^{} \leq -\varepsilon \} \;, 
\end{align*}
which are illustrated by different colors in Fig.~\ref{fig:Example_1}. The optimal control inputs can be computed by solving quadratic programs for different combinations of convex sets. In order to compute optimal control inputs, it is necessary to solve $5_{}^{N-1}$ convex quadratic programming problems.
\end{example}

\begin{example}
Let us consider the second-order system:
\begin{align}
	\x(t+1) = \begin{bmatrix} 0.9 & 0.2 \\ 0.8 & 1.5 \end{bmatrix} \x(t) + \begin{bmatrix} 0.6 \\ 0.8 \end{bmatrix} \bu(t) \;,
	\label{eqn:example_system}
\end{align}
with the initial condition $\x_{0}^{}=\left[\begin{smallmatrix} 0 & -1 \end{smallmatrix}\right]_{}^{\intercal}$. The event-triggered controller transmits control messages if the $\ell_{\infty}^{}$-norm of the state variable $\x(t)$ is greater than $\varepsilon = 0.25$. The performance indices are given by $Q=\mbox{diag}\{2,2\}$ and $R=5$. The control horizon is chosen as $N=7$. The state trajectories, depicted in Fig.~\ref{fig:Example_1}, can be obtained via solving the optimization problem~\eqref{eqn:opt_problem_cvx} for given sets $\Sigma_{1}^{}=\{4,4,4,1,1,0,0\}$, $\Sigma_{2}^{}=\{4,1,1,2,2,0,0\}$, and $\Sigma_{3}^{}=\{4,1,2,2,2,0,0\}$. Fig.~\ref{fig:Example_1} shows the state trajectories for two different sets. The set $\Sigma_{1}^{}$ leads to the global optimal solution. 
\end{example}

Example~1 demonstrates that obtaining the optimal solution to the finite-horizon optimal control problem, mentioned earlier, requires solving an exponential number of QPs. Therefore, in the next section, we propose heuristic algorithms that significantly reduce computational complexity, by compromising the optimality guarantees slightly.

\section{Computation of sub-optimal control actions} \label{sec:subopt_control}

In this section, we propose two heuristic algorithms that usually achieve a low control loss while significantly decreasing the computation time. We, firstly, develop a simple algorithm that takes the initial state $\x_0$ as input and finds a suitable sequence $\Sigma$ in a greedy manner by increasing the horizon of the problem from $1$ to $N$. Secondly, we apply an algorithm based on ADMM to compute the heuristic scheduling sequence $\Sigma$. This sequence is then used to solve the disjunctive problem \eqref{eqn:opt_problem_cvx} and obtain a good approximate control sequence $\pi$. Both techniques are of polynomial time complexity, hence, reducing the computational efforts significantly. Numerical investigations (see Section~\ref{sec:numerical}) indicate that these heuristic algorithms usually provide close to optimal solutions.

% --------------------------------------------------------------------------------------------------------------------------
%
% Greedy Algorithm
%
% --------------------------------------------------------------------------------------------------------------------------
\subsection{Greedy heuristic method}
We construct an intuitive greedy algorithm to solve the optimal control problem, proposed in~\eqref{eqn:opt_problem}, efficiently. As discussed earlier, given a feasible switching sequence $\Sigma$, a corresponding optimal control problem can be formulated as a quadratic programming problem of the form  \eqref{eqn:opt_problem_cvx}. One, then, solves the combinatorial problem~\eqref{eqn:opt_sequence_problem} to find the optimal switch sequence $\Sigma^\star$, thereby solving the optimal control problem \eqref{eqn:opt_problem}.

A good way to deal with the combinatorial nature of this type of algorithms and reduce the computational complexity is to employ a greedy algorithm, which solves a global problem by making a series of locally optimal decisions. Greedy algorithms do not necessarily provide optimal solutions; however, they might determine local optimal solutions, which can be used to estimate a globally optimal solution, in a reasonable time.

Our proposed greedy algorithm works as follows. Since $\x_{0}^{}$ is known, we can compute corresponding switching signal $\sigma(0)$. Then, we set $\Sigma = \{ \sigma(0), p \}$ for all $p \in\{0,1,\dots,2n\}$ and solve $2n+1$ number of QPs of the form~\eqref{eqn:opt_problem_cvx} to obtain the optimal switch signal $\sigma(1)$. The same procedure then repeats for $k=2,\dots, N-1$, where at each step $k$, we solve $2n+1$ times the QP problem~\eqref{eqn:opt_problem_cvx} to find the best $\sigma(k)$ keeping the previously found scheduling sequence $\Sigma = \{ \sigma(0), \dots, \sigma(k-1) \}$ unchanged.

Algorithm~\ref{alg:greedy_heuristic} provides the detailed steps of discussed greedy heuristic method. Algorithm~\ref{alg:greedy_heuristic} executes $(2n+1)(N-1)$ number of QPs compared to  $(2n+1)_{}^{N-1}$ as in optimal control procedure. As a result, it runs in polynomial time. 

%%%%%%%%%%%%%%%
%%%%%%%%%%%%%%%%
\begin{algorithm}[t]\label{alg:greedy_heuristic}
%\begin{algorithmic}
\SetAlgoLined
\KwData{  $\x_{0}$, $n$, $N$\;}
Compute $\sigma(0)$ based on $\x_{0}$\;
\For{Iteration count $k=2, \ldots, N$}{
 Obtain $\tilde{\x}=\x^\star$ and $\tilde{\bu}=\bu^\star$ by solving \eqref{eqn:opt_problem_cvx} with $\Sigma = \{ \sigma(0), \cdots, \sigma(k-2), p \}$ for all $p \in\{0,1,\cdots,2n\}$\;
 Compute $p_{}^{\star} = \mbox{argmin}_{\Sigma}^{}J_{\Sigma}^{\star}(\x(0))$\;
 Set $\sigma(k-1) = p_{}^{\star}$\;
 }
return $\Sigma$, $\tilde{\x}$, and $\tilde{\bu}$\;
%\end{algorithmic}
\caption{Greedy heuristic} % for solving problem \ref{e-prob}}
\end{algorithm}

% --------------------------------------------------------------------------------------------------------------------------
%
% ADMM
%
% --------------------------------------------------------------------------------------------------------------------------
\subsection{A heuristic method based on ADMM }\label{sec:ADMM}
In this section, we develop a computationally efficient method for solving~\eqref{eqn:opt_problem}. Our approach is based on, Alternating Direction Method of Multipliers (ADMM), a well developed technique to solve large-scale disciplined problems~\cite{BPC+:11}. The idea of utilizing ADMM as a heuristic to solve non-convex problems has been recently considered in literature~\cite{BPC+:11, TMB+:16, DTB:16, DBE+:13}. In particular, \cite{DTB:16, TMB+:16} employ ADMM to approximately solve problems with convex costs and non-convex constraints. The approximate ADMM solutions are  improved by multiplicity of random initial starts and a number of local search methods applied to ADMM solutions. 

In this paper, using some of the techniques introduced above, we first apply ADMM to original non-convex problem to achieve an intermediate solution. This algorithm, for a general non-convex problem, does not necessarily converge; however, when it does, the corresponding solution gives a good initial guess for finding the optimal solution to this problem. In a second phase of our heuristic method, we perform a polishing technique with the aim of achieving a feasible and hopefully closer to the optimal solution (see \cite{DTB:16} for an overview of polishing techniques). Our polishing idea is based on disjunctive programming problem~\eqref{eqn:opt_problem_cvx} with the fixed event index $\mathcal{T}$ and trajectory sequence $\Sigma$.

Initially developed to solve large-scale convex structured problems, ADMM has been recently advocated, to a great extent, for being effective approximation technique to solve non-convex optimization problems (see~\cite{BPC+:11, TMB+:16, DTB:16} and references therein). We start with casting~\eqref{eqn:opt_problem} to the ADMM form. Essentially, we have
\begin{equation}\label{eqn:opt_problem_ADMM}
\begin{array}{ll}
\underset{\z}{\mbox{minimize}}& \dfrac{1}{2}\z^\top F \z + \mathcal{I}(\w)\\
& G \z = \h,\\
& \z = \y,
\end{array}
\end{equation}
where $\z$ collects the state and control components; i.e., ${\z\triangleq [\x(0)^\top,\dots, \x(N)^\top,\bu(0)^\top, \dots, \bu(N-1)^\top]^\top}$, $F$ represents the positive definite Hessian of the quadratic cost. That is $F\triangleq \mbox{blkdiag}(Q_s,R_s)$ with $Q_s\triangleq I_{N+1}\otimes Q$ and $R_s\triangleq I_N \otimes R$. Moreover, $\mathcal{I}(\y)$ is the indicator function enforcing the event-triggering control law; that is
 \begin{equation}
 \label{eqn:opt_ADMM_indicator}
 \mathcal{I}(\y)=\left\{
 \begin{array}{ll}
\infty & \mbox{if}\; \eqref{eqn:opt_ADMM_violation} \;\mbox{holds},\\
0& \mbox{otherwise},
 \end{array}
 \right .
 \end{equation}
\begin{align}
\label{eqn:opt_ADMM_violation}
\{\exists t\in \{0, \dots, N-1\} \vert \y(t+1)\in \mathcal{C}_0\;  \mbox{and} \;\y(t+N+2)\neq 0_m\},
\end{align}
where \eqref{eqn:opt_ADMM_violation} detects the constraint violation in~\eqref{eqn:opt_problem} in terms of having $\x(t)\in \mathcal{C}_0$ while corresponding $\bu(t)\neq \mathbf{0}_m$. The constraints in~\eqref{eqn:opt_problem_ADMM} are twofold. The first constraint in~\eqref{eqn:opt_problem_ADMM} describes the LTI system dynamics~\eqref{eqn:Discrete_time_system}. Here $G\in \mathbb{R}^{nN \times \left(n(N+1)+mN\right)}$ have the following $i=\{1,\dots,N\},\; j=\{1,\dots 2N+1\}$- blocks,
\begin{equation}\nonumber
G_{ij} = \left\{
\begin{array}{ll}
A^{i-1}& \;\mbox{for } i=1\dots N,  j=1,\\
-I_n & \;j=i \;\mbox{and}\; j>1,\\
A^{i+N-j}B & \; \mbox{if}\; j\geq N+2\; \mbox{and}\; i+N\geq j,\\
0 & \; \mbox{otherwise}.
\end{array}
\right.
\end{equation}
and $\h\in\mathbb{R}^{n(N+1)+mN}\triangleq [\x(0)^\top, \mathbf{0}_n^\top,\dots, \mathbf{0}_m^\top]^\top$. Finally, the last constraint in~\eqref{eqn:opt_problem_ADMM} is of consensus-type to ensure that non-convex constraint in~\eqref{eqn:opt_problem} is asymptotically satisfied. 
 
 After formulating the problem, each iteration of ADMM algorithm consists of~(see \cite{BPC+:11} for a complete treatment of the subject):
 \begin{equation}\label{eqn:ADMM_algorithm}
 \begin{aligned}
 \z^{k+1/2} &= \underset{\z}{\mbox{argmin}} \left( \dfrac{1}{2}\z^\top F \z\right. \\
 &\left. +\dfrac{\rho}{2}\left\Vert  \left[ \begin{array}{l} G \\I\end{array}\right] \z  -\left[ \begin{array}{l} 0 \\I\end{array}\right] \z^k-\left[ \begin{array}{l} \h \\0\end{array}\right]+ \bu^k\right\Vert^2 \right) 
 \\
 \z^{k+1} &= \Pi (\z^{k+1/2}+\left[ \begin{array}{ll} 0& I \end{array}\right] \bu^k)\\
\bu^{k+1}&= \bu^k + \left[\begin{array}{l} G\\ I\end{array}\right] \z^{k+1/2}-\left[ \begin{array}{l} 0 \\I\end{array}\right] \z^k-\left[ \begin{array}{l} \h \\ 0\end{array}\right].
 \end{aligned}
 \end{equation}
Here, $k$ denotes the iteration counter, $\rho>0$ is the step-size (penalty) parameter, and $\Pi$ denotes the projection onto non-convex constraint and is given by
 \begin{equation}
 \Pi_i(\x) = \left\{\begin{array}{ll}
 0& \mbox{if}\; i\geq N+2 \mbox{ and } \x(i-N-1)\in \mathcal{C}_{0}^{}\\
 \x(i)& \mbox{otherwise}.
 \end{array}\right.
 \end{equation}
When the cost function in hand is convex and the constraints set is closed and convex then the ADMM algorithm converges to the  optimal solution of the problem. In our case, since $\mathcal{I}(\y)$ is non-convex the ADMM algorithm~\eqref{eqn:ADMM_algorithm} may not converge to optimal point. Moreover, examples can be found in which~\eqref{eqn:ADMM_algorithm} does not even converge into a feasible point. To further improve the ADMM solution, we utilize an additional convex problem to polish the intermediate results. That is, we restrict the search space to a convex set that includes the ADMM solution pattern. Let ${\tilde{\z}\triangleq [\tilde{\x}(0)^\top,\dots, \tilde{\x}(N)^\top,\tilde{\bu}(0)^\top, \dots, \tilde{\bu}(N-1)^\top]^\top}$  be the output of the ADMM algorithm and the sets $\widetilde{\Sigma}$ and {${\widetilde{\mathcal{T}}\triangleq\{t\in\mathbb{N}_{0}^{}; t<N : \tilde{\x}(t)\in\mathcal{C}_{0}^{}\}}$} denote the trajectory sequence associated with {$[\tilde{\x}_1^\top, \dots,  \tilde{\x}_N^\top]^\top$} and the restriction index of $\tilde{\x}$ into $\mathcal{C}_0$, respectively. Then our heuristic polishing procedure formulates~\eqref{eqn:opt_problem_cvx} with $\mathcal{T}=\mathcal{\widetilde{T}}$ and $\Sigma=\widetilde{\Sigma}$.  Algorithm~\ref{alg:ADMM_heuristic} provides a summary of our heuristic method. The input variables include the error-tolerance threshold $\epsilon^{tol}$, randomized initial state $\z^0$ as well as $\z_{best}$ and $f_{best}$ to store the final solution.
 %
 \iffalse
 the following problem
 \begin{equation}
 \label{eqn:ADMM_disjunctive}
 \begin{array}{ll}
\underset{z}{\mbox{minimize}}& \dfrac{1}{2}z^\top P z, \\
&\left[ \begin{array}{l}
F\\G
\end{array}\right] z = \left[ \begin{array}{l}
b\\0_{\vert \widetilde{\mathcal{T}}\vert m}
\end{array}\right],
\end{array}
 \end{equation}
 where $G\in\{0,1\}^{\vert\widetilde{\mathcal{T}}\vert m \times  \left(n(N+1)+mN\right) } $ and constructed such that for $t\in \widetilde{\mathcal{T}}$ it has the following block structure
 \begin{equation}\nonumber
 [0_{m\times (n(N+1)+(t-1)m)}, I_m, 0_{m\times (N-t)m }].
 \end{equation}
 \fi
 %

A few comments related to Algorithm~\ref{alg:ADMM_heuristic} are in order. 

 \begin{enumerate}
\item \EG{In contrast to the ADMM algorithms for convex problems that converge to the optimum for all positive range of the step-size parameter $\rho$, the stability of the ADMM algorithm for non-convex problems is sensitive to the choice of $\rho$.  Moreover, similar to the convex case, the convergence speed is also affected by the choice of $\rho$.}  A variety of techniques including proper step-size selection,  over relaxation, constraint matrix pre-conditioning and caching can be employed to accelerate the ADMM procedure~\eqref{eqn:ADMM_algorithm} (see \cite{ GTS+:15,GiB:16} for a reference on the topic).
\item One can utilize an adaptive iteration count procedure to improve the probability of finding feasible solution to~\eqref{eqn:opt_problem}. The procedure works as the following. Run the ADMM algorithm for a fixed number of iterations and then check if the disjunctive QP problem has feasible solution. If not, then increase the  number of iterations (e.g., double it) and repeat the procedure (run ADMM and disjunctive QP) until finding a feasible solution or reaching to a point where the current ADMM solution does not improve the previous one in terms of constraint violations and quadratic cost. 
\EG{\item The variable $z^0$ is initialized randomly using a normal distribution. In general, repeating the ADMM algorithm for a multiple of initial points may improve the approximate solution at the cost of extra computational complexity (as suggested in e.g.,~\cite{TMB+:16}). However, in our application  we did not experience significant performance improvement by repeating the ADMM algorithm (steps $1-6$ in Algorithm~\ref{alg:ADMM_heuristic}) with different initializations. One reason could be that in Algorithm \ref{alg:ADMM_heuristic}, we employ the solution trajectories $\widetilde{\Sigma}$ and $\widetilde{\mathcal{T}}$ based on the  ADMM solution $\tilde{z}$ and not $\tilde{z}$ itself in the disjunctive problem~\eqref{eqn:opt_problem_cvx}.}
\item Using caching and $LDL^\top$ matrix factorization techniques in the first-step of~\eqref{eqn:ADMM_algorithm} \cite{BPC+:11}, each $\z$-update in Algorithm~\ref{alg:ADMM_heuristic} costs on the order of $O\left((n(N+1)+mN)^2\right)$ for dense $P$ matrices. Furthermore, if we assume $N\geq \max\{n,m\}$ then each ADMM iteration costs on the order of $O(N^2)$. This means that overall cost of the ADMM algorithm is of $O(K N^2)$ where $K$ is the number of ADMM iterations. In our application, with proper algorithm parameter selection, the ADMM algorithm converges almost always within a few hundreds  of iterations  (see the results presented in Section~\ref{sec:numerical}).
\item The computational complexity of the disjunctive step in our heuristic method falls into the generic complexity of convex QPs  on the order of $O(\left(\vert \mathcal{T} \vert 2n +(N-\vert \mathcal{T}\vert)(2n-1)\right)^2)$ \cite{BoV:04}.
\end{enumerate}
 
%%%%%%%%%%%%%%%
%%%%%%%%%%%%%%%%
\begin{algorithm}[t]\label{alg:ADMM_heuristic}
\SetAlgoLined
\KwData{  $\epsilon^{tol}$, $\z^0\sim\mathcal N(0,\sigma^2I)$, $\z_{best}=\emptyset$, and $f_{best}=\infty$\;}
\For{Iteration count $k=1, 2, \ldots, K$}{
 update $\z$ from~\eqref{eqn:ADMM_algorithm}\;
 
\If{$\Vert G\z^k-\h\Vert\leq \epsilon^{tol}$ \mbox{and} $(1/2)\z^{k\top} F \z^k < f_{best}$}{
 $\z_{best}= \z^k$\;
 }
 }
 \If{$f_{best} < \infty$}{Obtain $\z_{best}$ from disjunctive problem~\eqref{eqn:opt_problem_cvx}\;
}
return $\z_{best}$\;
\caption{ADMM-Disjunctive heuristic} 
\end{algorithm}

% --------------------------------------------------------------------------------------------------------------------------
%
% Receding Horizon Control
%
% --------------------------------------------------------------------------------------------------------------------------

\section{Receding horizon control}

Except a few special cases, finding a solution to an infinite horizon optimal control problem is not possible. Receding horizon control is an alternative scheme to infinite horizon problem that repeatedly requires solving a constrained optimization problem. At each time instant $t\geq 0$, starting at the current state $\x(t)$ (assuming that a full measurement of the state $\x(t)$ is available at the current time $t\in\mathbb{N}_{0}^{}$), the following cost function 
\begin{multline*}
	J(\x(t),\boldsymbol{\pi}_{t}^{\star N-1}) = V_{f}^{}(\x(t+N\mid t)) \\ + \sum_{k=t}^{t+N-1}\ell(\x(k\mid t),\bu(k\mid t))
\end{multline*}
subject to system dynamics and constraints involving states and controls is minimized over a finite horizon $N$. Here, the function $\ell$ defines the stage cost, and $V_{f}^{}$ defines the terminal cost. Denote the minimizing control sequence, which is a function of the current state $\x(t)$, by
\begin{align*}
	\boldsymbol{\pi}_{t}^{\star N-1}=\big[ \bu^{\star}(t\mid t)^{\intercal}, \cdots, \bu^{\star}(t+N-1\mid t)^{\intercal} \big] \;,
\end{align*} 
then the control applied to the plant at time $t\geq 0$ is the first element of this sequence, that is, 
\begin{align*}
	\bu(t) = \big[ I ~ 0 ~ \cdots ~ 0 \big] \boldsymbol{\pi}_{t}^{\star N-1} \;. \label{eqn:receding_horizon_strategy}
\end{align*}
Time is then stepped forward one instant, and the procedure described above is repeated for another $N$-step ahead optimization horizon. 

\begin{figure}[!t]\centering
	\includegraphics[scale=0.9]{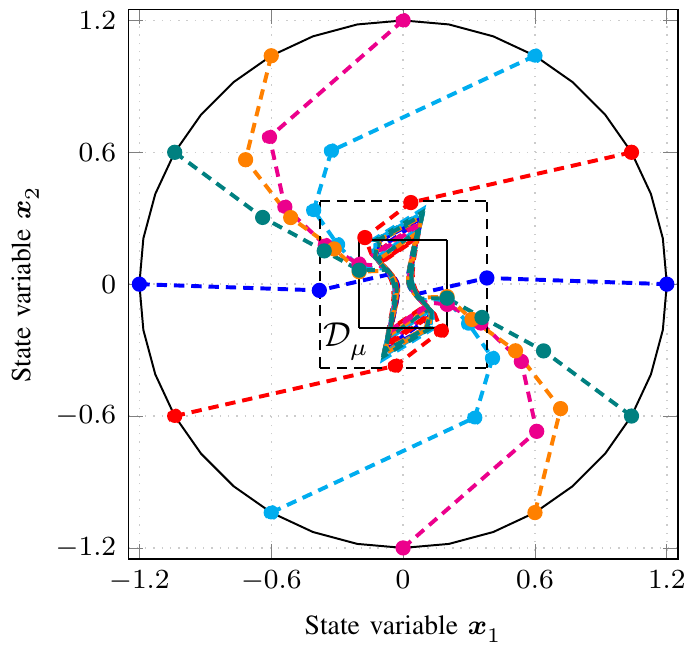}
  \caption{Evolution of the state trajectories of the system~\eqref{eqn:example_system} under the receding horizon strategy~\eqref{eqn:receding_horizon_strategy} and constraints~\eqref{eqn:event_triggered_set}--\eqref{eqn:event_triggered_constraint_time} for a set of initial conditions $\x_{0}^{}$.}
  \label{fig:RHC_example} 
\end{figure}

It is worth noting that, due to the fundamental limitation of the optimization problem~\eqref{eqn:opt_problem}, the state $\x(t)$ never converges to the origin if the system is unstable. It can only converge into a set including the origin. The following result shows the existence of this set:

\begin{theorem}\label{thm:Stability}
	Suppose that $\mathcal{D}_{\mu}^{} \triangleq \{ \x\in\mathbb{R}^{n} \,:\, \parallel \x \parallel_{\infty}^{} \leq \mu \}$ is a neighborhood of the origin, where
	\begin{align}
		\mu &\triangleq \sqrt{\frac{\kappa\lambda_{\max}(A^{\intercal}PA+Q) n \varepsilon^{2}}{\lambda_{\min}^{2}(Q)}} 
	\end{align}
	for some $\kappa > 0$. Then, the system~\eqref{eqn:Discrete_time_system} with event-triggering constraints~\eqref{eqn:event_triggered_constraint} and \eqref{eqn:event_triggered_constraint_time} is uniformly practically asymptotically stable, i.e., $\displaystyle{\lim_{t\rightarrow\infty} \parallel \x(t) \parallel_{\infty}^{} \leq \mu}$.
\end{theorem}

Theorem~\ref{thm:Stability} establishes practical stability of the system~\eqref{eqn:Discrete_time_system} with event-triggering constraints \eqref{eqn:event_triggered_constraint} and~\eqref{eqn:event_triggered_constraint_time}. It shows that if provided conditions are met, then the plant state will be ultimately bounded in an $\ell_{\infty}^{}$-norm ball of radius $\mu$. It is worth noting that, as the other stability results, which use Lyapunov techniques, this bound will not be tight. Next, we would like to exemplify the convergence of the state $\x(t)$ to the set $\mathcal{D}_{\mu}^{}$ from a set of initial conditions $\x_{0}^{}$.

\begin{example}
	Consider the second-order system, given by~\eqref{eqn:example_system}, with initial conditions $\x_{0}^{}=1.2\left[\begin{smallmatrix} \text{sin}(\frac{\pi k}{6}) ~ \text{cos}(\frac{\pi k}{6}) \end{smallmatrix}\right]^{\intercal}$ for all $0 \leq k \leq 12$. The controller transmits a control message whenever the event-triggering condition, i.e., $\parallel\x(t)\parallel_{\infty}^{} > 0.25$, holds. The performance indices are chosen as $Q=\mbox{diag}\{2,2\}$ and $R=5$ while the prediction horizon is set to $N=6$. Fig.~\ref{fig:RHC_example} shows the state trajectories of the receding horizon implementation of the event-triggered control system, described by \eqref{eqn:Discrete_time_system} -- \eqref{eqn:event_triggered_constraint_time}, for different initial conditions $\x_{0}^{}$. As can be seen in the same figure, the number of transmitted control commands depends on the initial condition. 
\end{example}

%====================================================================================================
%
% NUMERICAL EXAMPLES
%
%====================================================================================================
\section{Numerical examples}\label{sec:numerical}

\subsection{Performance of sub-optimal event-triggered control actions}
To illustrate performance benefits of the proposed heuristic algorithms, we consider a third-order plant with the following state-space representation:
\BD{
\begin{multline}
	\x(t+1) = \begin{bmatrix} 0.53 & -2.17 & 0.62 \\ 0.22 & -0.06 & 0.51 \\ -0.92 & -1.01 & 1.69 \end{bmatrix} \x(t) 
	+ \begin{bmatrix} 0.4 \\ 0.7 \\ 0.9 \end{bmatrix}\bu(t) \\ + B_{\w}^{}\w(t) \;. \label{eqn:example_3d}
\end{multline}
}
Assume that there is no disturbance acting on the plant, i.e., $B_{\w}^{} = 0$.

Our aim, here, is to minimize~\eqref{eqn:cost_function} with performance indices $Q_{}^{} = P_{}^{} = \mathrm{diag}\{2,2,2\}$ and $R_{}^{} = 5$, and the horizon length $N = 8$. The initial condition is chosen as
\begin{align*}
	\x_{0} = 
	\begin{bmatrix}
		\sin\theta\cos\phi & \sin\theta\sin\phi & \cos\theta
	\end{bmatrix}^{\intercal} \;,
\end{align*}
where $0 \leq \theta \leq \pi$ and $0 \leq \phi \leq \pi$. To perform simulations, we selected $577$ different data points, which are equidistant over a half spherical surface. We evaluated the performance of the greedy and the ADMM-based heuristic algorithms for all these initial conditions. For each data point on the half sphere, we run $823,543$ number of QPs to obtain the global optimal solution. In total, we run more than $475$ million number of QPs in our evaluation. We used a linux machine with $32$ cores in Intel Core $i7$ Extreme Edition $980X$ to solve these QPs in parallel threads. Within this computing resources, it takes approximately $6$ days to perform simulations for one problem setup.

To compare the (true) optimal control performance with our heuristic method, we performed Algorithm~\ref{alg:ADMM_heuristic} on the same problem set. In particular, we evaluated three sets of problem with $\varepsilon = 0.2$, $\varepsilon=0.4$, and ${\varepsilon=0.6}$. It is generally expected that the control problem becomes more challenging with increasing the event-threshold $\varepsilon$. We run the ADMM-based heuristic algorithm with adaptive iteration count procedure described in Section \ref{sec:ADMM}. Moreover, we picked $\rho=9.8$, $\rho=5.8$, and $\rho=6.9$ respectively, for the aforementioned problem scenarios. Under this setup, in both cases (i.e., $\varepsilon = 0.2$ and $\varepsilon=0.4$), the ADMM method produced always feasible solutions for maximum $300$ number of ADMM-iterations. However, when $\varepsilon = 0.6$ and $\rho=6.9$, the method created infeasible solutions for two initial conditions: $\x_{0}^{}=\left[ 0.3036,\, 0.2330,\, 0.9239 \right]^\top$ and $\x_{0}^{}=\left[0.6830, \, 0.1830, \, 0.7071 \right]^\top$. To compute feasible solutions, we needed to adjust $\rho$ accordingly for these initial conditions. 

\begin{figure}[!t]\centering
	\includegraphics[scale=0.55]{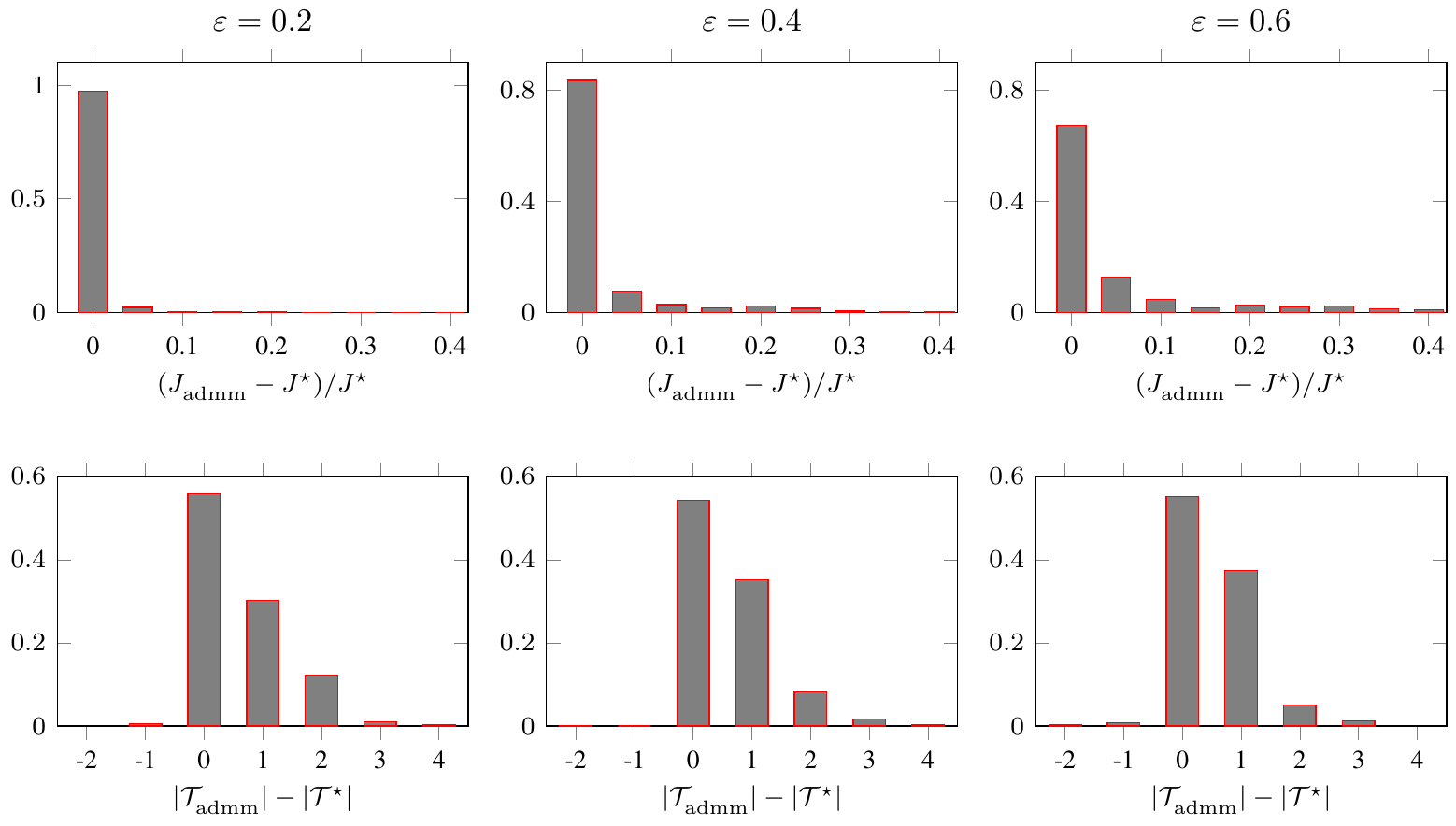}
  \caption{Relative error $(J_{\mathrm{admm}}^{}-J_{}^{\star})/J_{}^{\star}$ and control-action cardinality difference of the (sub-optimal) ADMM-based heuristic solution from the optimal cost for $577$ problem instances of third order plant model~\eqref{eqn:example_3d}.
  }
  \label{fig:comparison_admm_vs_optimal} 
\end{figure}

Fig.~\ref{fig:comparison_admm_vs_optimal} shows the outcome of the comparison. As it shows, for $\varepsilon=0.2$, the ADMM-based heuristic algorithm is able to find near optimal solutions (with relative error less than $5\%$) in almost all cases, and for $\varepsilon=0.4$ and $\varepsilon=0.6$, the  algorithm finds solutions with relative error less than {$5\%$} in more than $80\%$ of problem instances. The average optimality gap $(J_{\text{admm}}-J^\star)/J^\star$ for the three cases were $0.0035$, $0.0237$, and $0.0716$, respectively. Finally, by comparing the cardinality of control event index $\mathcal{T}$ for the optimal solutions and our ADMM method, one concludes that the heuristic method in more than $50\%$ of cases finds similar-sparse solution as the optimum ones, and in the rest of cases it tends to find more sparse control actions. 

We also compared the true optimal control performance with a greedy heuristic method, shown in Algorithm~\ref{alg:greedy_heuristic}, for three sets of problems, mentioned above, with $\varepsilon = 0.2$, $\varepsilon=0.4$, and ${\varepsilon=0.6}$. The greedy algorithm detected $22$, $90$, and $135$ optimal solutions of the aforementioned $577$ problems, respectively. Also, it computed the optimal solution with relative error less than $5\%$ in more than $65\%$ of problems at hand for all three problem setups. It is important to note that the greedy heuristic method always provided feasible solutions for all problem cases. The average optimality gaps $(J_{\text{greedy}}-J^\star)/J^\star$ in all three cases were $0.0514$, $0.0780$, and $0.0826$, respectively.  As compared to the ADMM-based heuristic method, the optimality gaps became larger.

\begin{figure}[!t]\centering
	\includegraphics[scale=0.55]{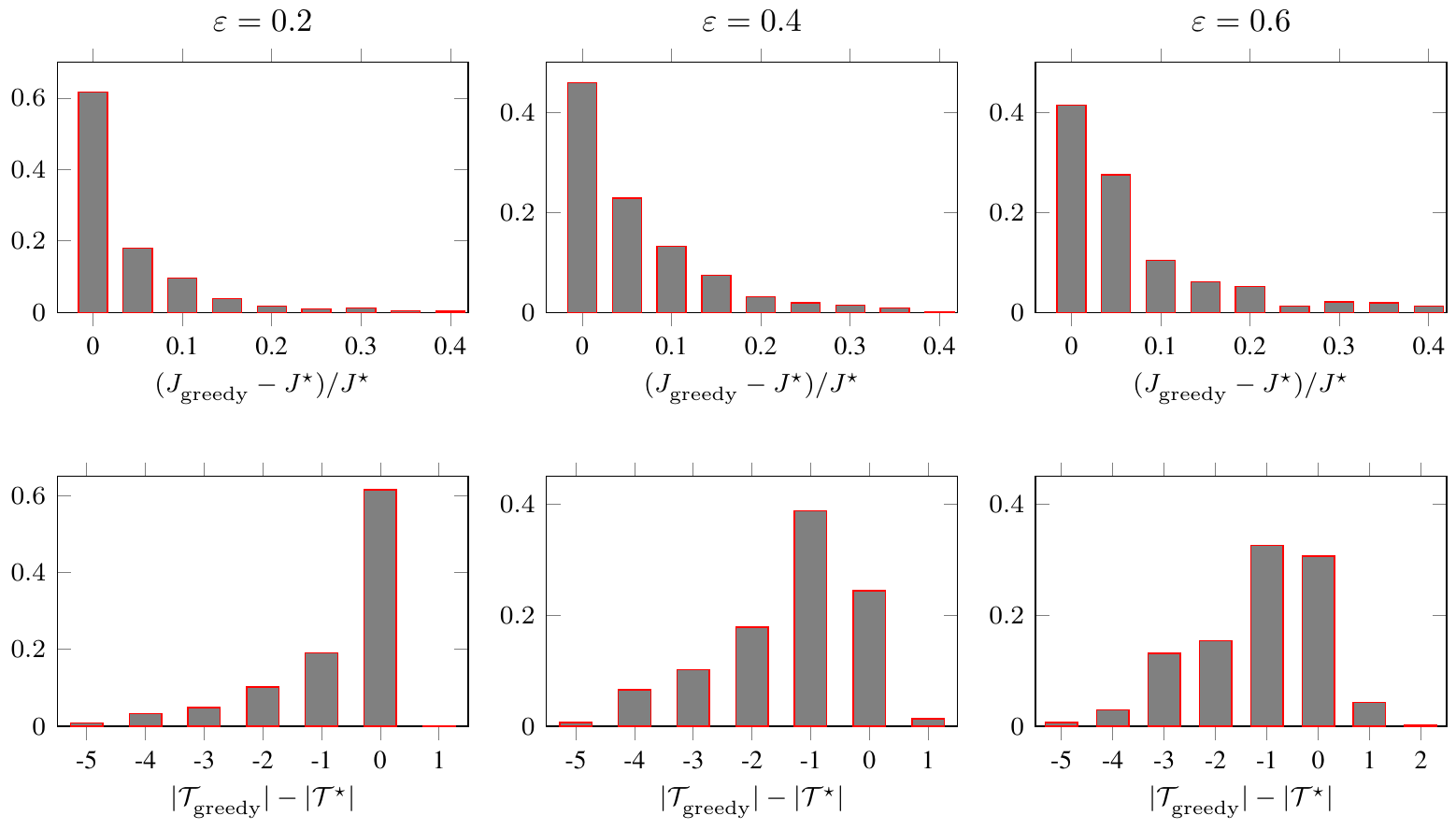}
  \caption{Relative error $(J_{\mathrm{greedy}}^{}-J_{}^{\star})/J_{}^{\star}$ and control-action cardinality difference of the greedy solution from the optimal cost for $577$ problem instances of third order plant model~\eqref{eqn:example_3d}.
  }
  \label{fig:comparison_greedy_vs_optimal} 
\end{figure}
\begin{figure}[!t]\centering
	\includegraphics[scale=0.6]{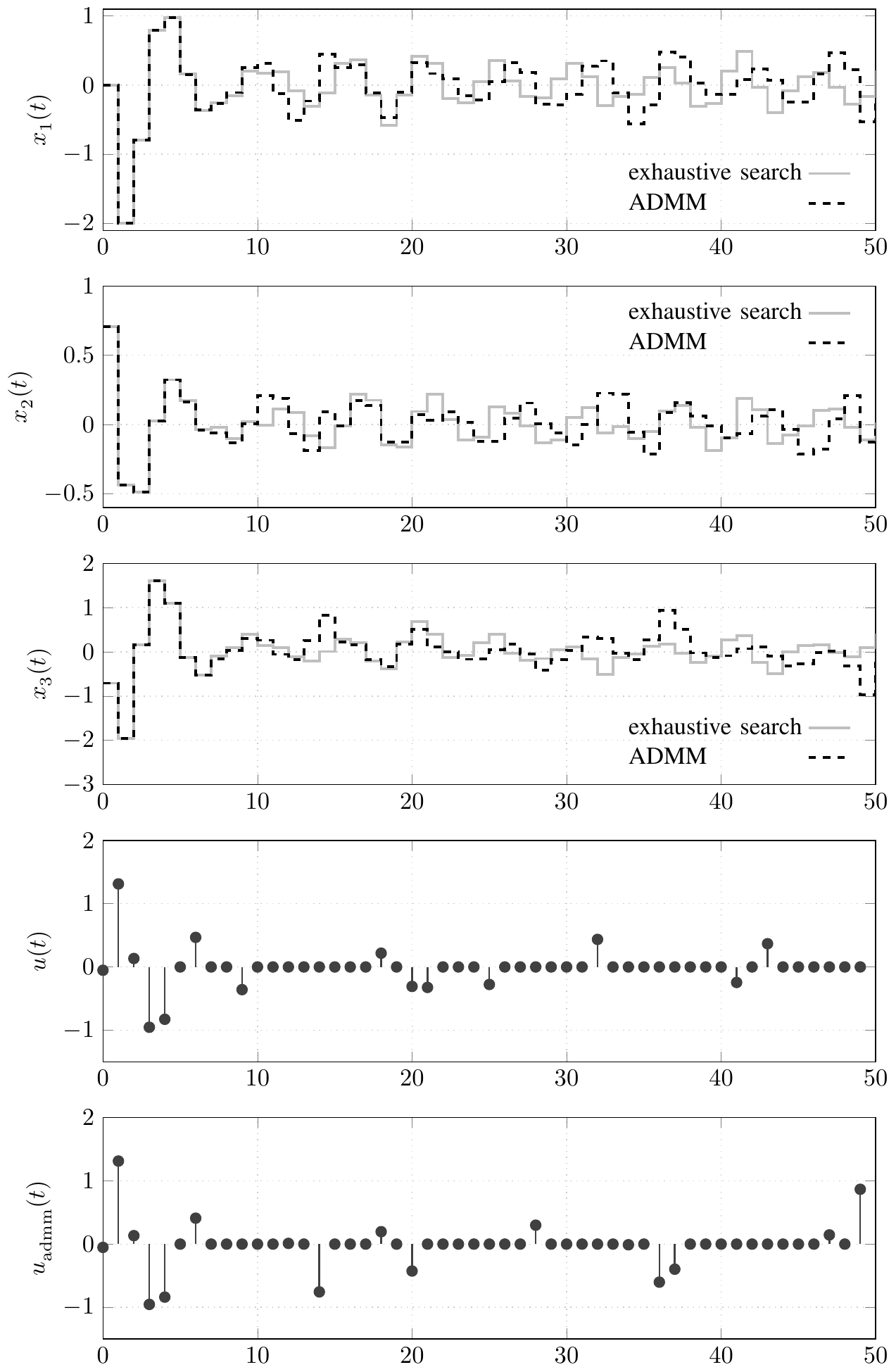}
  \caption{\BD{States and input versus time for the event-triggered model predictive controller presented in Section~\ref{sec:numerical_mpc}. In this figure, $\x_{1}^{}(t)$, $\x_{2}^{}(t)$ and $\x_{3}^{}(t)$ represent state variables whereas $u(t)$ and $u_{\mathrm{admm}}^{}(t)$ denote the optimal and the heuristic control inputs computed based on the exhaustive search and the ADMM algorithm, respectively.}
  }
  \label{fig:receding_horizon_control} 
\end{figure}

\BD{
\subsection{Receding horizon control implementation}\label{sec:numerical_mpc}
We consider the open-loop unstable discrete-time system represented by~\eqref{eqn:example_3d} with $B_{\w}^{} = 0$. The prediction horizon is chosen to be $N = 6$ with performance indices $Q_{}^{} = P_{}^{} = \mathrm{diag}\{2,2,2\}$ and $R_{}^{} = 5$. The event-threshold $\varepsilon$ for the event-triggered receding horizon implementation is set to $0.4$. Fig.~\ref{fig:receding_horizon_control} demonstrates the time responses of the event-triggered receding horizon control system for the initial condition $\x_{0}^{}=\left[0, \, \frac{\sqrt{2}}{2}, \, -\frac{\sqrt{2}}{2} \right]^\top$. Here, two different approaches for computing control input trajectories are considered: the exhaustive search algorithm and the heuristic ADMM algorithm with $\rho=4.8$. As can be seen in Fig.~\ref{fig:receding_horizon_control}, while the state trajectories associated to two control approaches deviate from one another (except for the few first sampling instants that they match together), the corresponding control inputs follow a rather similar sparsity pattern. The event-triggered receding horizon control via exhaustive search algorithm uses the communication channel $14$ times over the $50$ sampling instants, which results in a reduction of $72\%$ in communication and computational burden, whereas the event-triggered receding horizon control via ADMM algorithm uses the channel $16$ times over the $50$ sampling instants, which leads to a reduction of $68\%$ in communication and computational burden. For the aforementioned initial value, the exhaustive search achieves a better control cost: $65.42$ compared to $77.72$ as of the control cost of the ADMM-based heuristic. However, this conclusion does not hold in general. In particular, we experienced cases where starting from a different initial value, the receding horizon controller derived by exhaustive search performs worse than the one based on ADMM-based heuristic.
}

\subsection{Trade-off between communication rate and control performance}
We consider the case where a random disturbance $\w(t)\in\mathbb{R}_{}^{n}$, which is modeled by a discrete-time zero-mean Gaussian white process with co-variance $\Sigma_{\w}^{}$, is acting on the plant characterized by the injection matrix $B_{\w}^{} = \mbox{diag}\{1,1,1\}$. The initial condition $\x_{0}^{}$ is modeled as a random variable having a normal distribution with zero mean and co-variance $\Sigma_{\x_{0}^{}}^{}$. The process noise $\w(t)$ is independent of the initial condition $\x_{0}^{}$.

The control performance is measured by a quadratic function:
\begin{align}
	J_{\infty}^{} = \frac{1}{500}\sum_{t=0}^{499} \big(\x_{}^{\intercal}(t)Q\x_{}^{}(t) + \bu_{}^{\intercal}(t)R\bu(t)\big)\;, \label{eqn:empirical_cost}
\end{align}
while the communication rate is measured by 
\begin{align}
	\pi_{\infty}^{} = \frac{1}{500}\sum_{t=0}^{499} \mathbbm{1}_{\{\parallel\x(t)\parallel_{\infty}^{}\geq 0\}}^{} \;. \label{eqn:empirical_rate}
\end{align}

\begin{figure}[!t]\centering
	\includegraphics[scale=0.7]{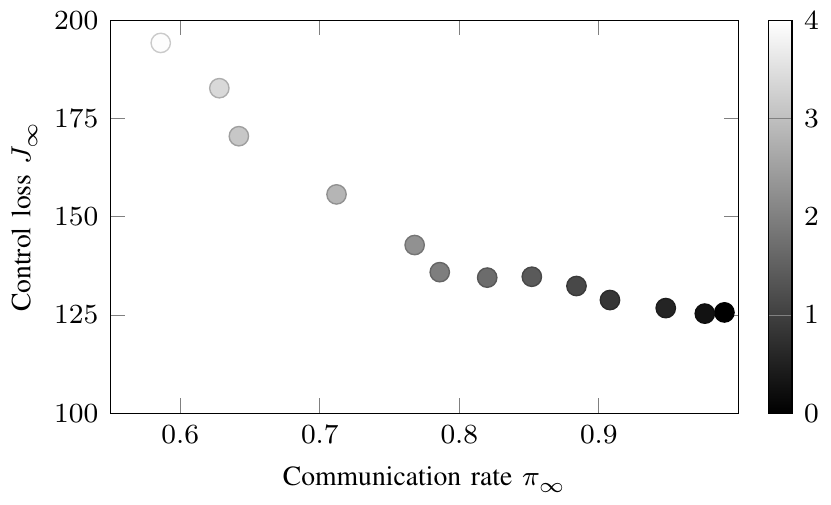}
  \caption{A comparison of the control performance and the communication frequency for different event-threshold values $\varepsilon > 0$ (shown in gray scale). The marked curve with is obtained by averaging $2,500$ Monte Carlo simulations for the horizon length $500$ samples with the process noise $\{\w(t)\}_{t\in\mathbb{N}_{0}^{}}$and the initial condition $\x_{0}^{}$ generated randomly.}
  \label{fig:tradeoff} 
\end{figure}

The visualization of the trade-off between the control loss and the communication rate is demonstrated in Fig.~\ref{fig:tradeoff}. Different data points on the curve correspond to different event-threshold values ranging from $0.5$ to $4.0$, and the control loss~\eqref{eqn:empirical_cost} and the communication rate~\eqref{eqn:empirical_rate} are evaluated via Monte Carlo simulations. Note that the communication rate decreases dramatically with an increased control loss as the threshold $\varepsilon$ varies between $0$ and $2.25$ (dark colors). For $\varepsilon > 2.25$ (lighter color), both quantities become less sensitive to changes in the threshold value. We can also identify $0\leq\varepsilon\leq 1.5$ as a particularly attractive region where a large decrease in the communication rate can be obtained for a small loss in control performance.

%

%====================================================================================================
%
% CONCLUSION
%
%====================================================================================================

\section{Conclusion}

In this paper, we considered a feedback control system where an event-triggering rule dictates the communication between the controller and the actuator. We investigated the optimal control problem with an additional constraint on the event-triggered communication. While this optimization problem, in general, is non-convex, we used a disjunctive programming formulation to obtain the optimal solution via solving an exponential number of quadratic programs. Then, we proposed a heuristic algorithms to reduce the computational efforts while achieving an acceptable performance. Later, we provided a complete stability analysis of receding horizon control that uses a finite-horizon optimization in the proposed class. Our numerical study confirmed the theory.

%====================================================================================================
%
% APPENDIX
%
%====================================================================================================

\section{Appendix}

We begin with the definition of the \textit{practical stability} of control systems.

\begin{definition}[practical stability~\cite{JiW:02}]
	The system~\eqref{eqn:Discrete_time_system} is said to be uniformly practically asymptotically stable in $\mathcal{A}\subseteq\mathbb{R}^{n}$ if $\mathcal{A}$ is a positively invariant set for~\eqref{eqn:Discrete_time_system} and if there exist a $\mathcal{KL}$-function $\beta$, and a nonnegative constant $\delta\geq 0$ such that
	\begin{align*}
		\parallel \x(t) \parallel \leq \beta(\parallel \x_{0} \parallel, t) + \delta \;.
	\end{align*} 
\end{definition}
%\EG{@Burak: I guess it is good to cite above definition, besides what is $\beta (\cdot,t)$?}

\begin{definition}[practical-Lyapunov function~\cite{JiW:02}]
	A function $V:\mathbb{R}^{n}\rightarrow\mathbb{R}_{\geq 0}$ is said to be a practical-Lyapunov function in $\mathcal{A}$ for the system~(1) if $\mathcal{A}$ is a positively invariant set and if there exist a compact set, $\Omega\subseteq\mathcal{A}$, neighbourhood of the origin, some $\mathcal{K}_{\infty}$-functions $\alpha_{1}, \alpha_{2}$ and $\alpha_{3}$, and some constants $d_{1}, d_{2}\geq 0$, such that 
	\begin{align}
		V( \x ) &\geq \alpha_{1}(\parallel \x \parallel), \forall\x\in\mathcal{A} \;, \\
		V( \x ) &\leq \alpha_{2}(\parallel \x \parallel) + d_{1}, \forall\x\in\Omega \;, \\
		V( f(\x) ) - V( \x ) &\leq -\alpha_{3}(\parallel \x \parallel) + d_{2}, \forall\x\in\mathcal{A} \;.
	\end{align}
\end{definition}
% \EG{@Burak: what is $f(x)$ in above definition?}
% \EG{@Burak:I could not find the exact definition in [34], would you please refer to the exact place within the reference. Like [34, Definition xxx]}
%~\cite{AgQ:13}

\noindent\textbf{Proof of Theorem~\ref{thm:Stability}.}
To prove the practical stability of the system~\eqref{eqn:Discrete_time_system} with a set of control actions $\boldsymbol{\pi}_{}^{\star} = \{ \bu^{\star}(t\mid t)^{}, \cdots, \bu^{\star}(t+N-1\mid t)^{} \}$, which fulfills constraints~\eqref{eqn:event_triggered_constraint} and~\eqref{eqn:event_triggered_constraint_time}, we will analyze the value function, 
\begin{align*}
	V_{N}^{\star}(\x(t)) \triangleq \min_{\boldsymbol{\pi}}^{}J(\x(t),\boldsymbol{\pi}) , 
\end{align*}
where $J(\x(t),\boldsymbol{\pi})$ is as seen in~\eqref{eqn:cost_function}. We borrow the shifted sequence approach, described in~\cite{RaM:09}, and we use a feasible control sequence, i.e., $\tilde{\boldsymbol{\pi}} = \{\bu_{}^{\star}(t+1 \mid t), \cdots, \bu_{}^{\star}(t+N-1 \mid t), \hat{\bu} \}$. By the optimality property, we obtain the following bound (with $\x_{}^{}(t \mid t) = \x(t)$ and $\bu_{}^{\star}(t \mid t) = \bu(t)$):
\begin{align}
	V_{N}^{\star}(\x(t+1)) & - V_{N}^{\star}(\x(t)) \leq V_{N}^{}(\x(t+1),\tilde{\boldsymbol{\pi}}) - V_{N}^{\star}(\x(t)) \nonumber \\
	& = - \ell(\x(t),\bu(t)) + V_{f}\big(A\x(t+N \mid t)+B\hat{\bu}\big) \nonumber\\
	& - V_{f}(\x(t+N \mid t)) + \ell(\x(t+N \mid t),\hat{\bu}) \label{eqn:optimality_bound}
\end{align}
for all $\x(t)\in\mathbb{R}^{n}$.
% Thus,~\eqref{eqn:optimality_bound} holds if the following one is satisfied:
%
We now investigate the quantity of
\begin{align*}
	\Delta V_{f}^{}(t) + \ell(\x(t+N \mid t),\hat{\bu})   \;,
\end{align*}
where $\Delta V_{f}^{}(t) \triangleq V_{f}\big(A\x(t+N \mid t)+B\hat{\bu}\big) - V_{f}(\x(t+N \mid t))$. 
Based on $\x(t+N \mid t)$, there are two possibilities: 
\begin{itemize}
	\item Suppose that $\parallel \x(t+N \mid t) \parallel_{\infty}^{} > \varepsilon$, then for any feasible control law, i.e., $\hat{\bu}=K\x(t+N \mid t)$, the following inequality holds:
	\begin{align}\label{eqn:deltaV_case_1}
		& \Delta V_{f}^{}(t) + \ell(\x(t+N \mid t),\hat{\bu})\nonumber \\
		& = \x(t+N \mid t)_{}^{\intercal} \big( A_{K}^{\intercal}PA_{K}^{} - P + Q_{}^{*} \big)\x(t+N \mid t)_{}^{} < 0 \;,
	\end{align}
	where $A_{K}^{} = A + BK$ (i.e., Schur) and $Q_{}^{*} = Q + K_{}^{\intercal}RK$. From~\eqref{eqn:optimality_bound} and \eqref{eqn:deltaV_case_1}, it follows that 
	\begin{equation}\nonumber
	V_{N}^{\star}(\x(t+1)) - V_{N}^{\star}(\x(t)) < 0 \;.
	\end{equation}
		\item Suppose that $\parallel \x(t+N \mid t) \parallel_{\infty}^{} \leq \varepsilon$, then $\hat{\bu} = 0$. Hence, it follows that:
	\begin{align*}
		\Delta V_{f}^{}(t) + & \ell(\x(t+N \mid t),\hat{\bu}) \\
		& = \x(t+N \mid t)_{}^{\intercal} \big( A^{\intercal}PA - P + Q \big)\x(t+N \mid t)_{}^{} \\ 
		& \leq \x(t+N \mid t)_{}^{\intercal} \big( A^{\intercal}PA + Q \big)\x(t+N \mid t)_{}^{} \\
		& \leq \lambda_{\max}^{}\big( A^{\intercal}PA + Q \big) \parallel \x(t+N \mid t)_{}^{} \parallel_{2}^{2} \\
		& \leq \lambda_{\max}^{}\big( A^{\intercal}PA + Q \big) n \parallel \x(t+N \mid t)_{}^{} \parallel_{\infty}^{2} \\
		& \leq \lambda_{\max}^{}\big( A^{\intercal}PA + Q \big) n \varepsilon{}^{2} \;, 
	\end{align*}
where we used the norm inequality $\parallel \v \parallel_{2}^{} \leq \sqrt{n} \parallel \v \parallel_{\infty}^{}$ for any $\v \in \mathbb{R}_{}^{n}$; see~\cite{Ber:09}.
\end{itemize}
As a result, we conclude that, for any $\x(t+N \mid t)_{}^{} \in \mathbb{R}^{n}$, the following inequality holds:
\begin{align}
%	V_{f}\big(A\x(t+N \mid t)+B\tilde{\bu}\big) - V_{f}(\x(t+N \mid t)) \\
		\Delta V_{f}(t) + \ell(\x(t+N \mid t),\hat{\bu}) \leq \eta\;,
		\label{eqn:Bound_on_terminal_cost}
\end{align}
where $\eta = \lambda_{\max}^{}\big( A^{\intercal}PA + Q \big) n \varepsilon{}^{2}$.
Inserting~\eqref{eqn:Bound_on_terminal_cost} and $a_{1}^{}\parallel \x(t) \parallel_{2}^{2} \leq \ell(\x(t),\bu(t))$ where $a_{1}^{}\triangleq\lambda_{\min}(Q)$ into~\eqref{eqn:optimality_bound} yields:
\begin{align}\label{eqn:Lyapunov_ineq2}
	V_{N}^{\star}(\x(t+1)) & - V_{N}^{\star}(\x(t)) \leq -\lambda_{\min}(Q) \parallel \x(t) \parallel_{2}^{2} + \eta \;.
\end{align}

Now, let the binary variable $\delta(t)\in\{0,1\}$, for each time instant $t$,  represent the transmission of the control action as follows:
\begin{align}
	\delta_{}^{}(t) = 
	\begin{cases}
		0 & \text{if} ~ \parallel \x(t) \parallel_{\infty}^{} \leq \varepsilon \;, \\
		1 & \text{otherwise} \;.
	\end{cases}
\end{align}
For any feasible control sequence, i.e., $\boldsymbol{\pi}_{}^{\prime} = \{ K \x(t), \cdots, K \x(t+N-1 \mid t) \}$, there exists an associated scheduling sequence, denoted by $\Delta_{}^{}(t) = \{\delta_{}^{}(t), \cdots, \delta_{}^{}(t+N-1)\}$. Applying these feasible control inputs, the upper bound of $V_{N}^{\star}(\x)$ becomes:
\begin{align}\label{eqn:V_upperbound}
	V_{N}^{\star}(\x(t)) &\leq \x^{\intercal}(t) S_{\Delta_{}^{}(t)}^{} \x(t) \leq \lambda_{\max}(S_{\Delta_{}^{}(t)}^{})\parallel \x(t) \parallel_{2}^{2} \;, \nonumber \\
	& \leq  a_{3}^{} \parallel \x(t) \parallel_{2}^{2}  \;,
\end{align}
where $a_{3}^{} \triangleq \displaystyle{\max_{\Delta\in\mathcal{S}}^{}\lambda_{\max}(S_{\Delta_{}^{}}^{})}$ and
\begin{align*}
	S_{\Delta_{}^{}}^{} = Q_{\delta(0)}^{} + \sum_{i=1}^{N-1}\prod_{j=0}^{i-1} A_{\delta(j)}^{\intercal}Q_{\delta(j)}^{}A_{\delta(j)}^{} + \prod_{k=0}^{N-1}A_{\delta(k)}^{\intercal}QA_{\delta(k)}^{} \;,
\end{align*}
with $Q_{\delta(i)}^{} \triangleq \delta(i)Q_{}^{*} + (1-\delta(i))Q_{}^{}$ and $A_{\delta(i)}^{} \triangleq \delta(i)A_{K}^{} + (1-\delta(i))A_{}^{}$ for all $i\in\{0, \cdots, N-1\}$. Here, $\mathcal{S}$ denotes a set of all possible feasible schedules generated by a stabilizing control sequence $\boldsymbol{\pi}_{}^{\prime}$.

To obtain the lower bound, we have $\x_{}^{\intercal}(t) Q \x(t) \leq V_{N}^{\star}(\x(t))$ and hence $a_{2}^{} \parallel \x(t) \parallel_{2}^{2} \leq V_{N}^{\star}(\x(t))$ where $a_{2}^{}\triangleq\lambda_{\min}^{}(Q)$. From, \eqref{eqn:Lyapunov_ineq2} and \eqref{eqn:V_upperbound} we establish the following relation:
\begin{align}
	V_{N}^{\star}(\x(t+1)) \leq \bigg( 1 - \frac{a_{2}^{}}{a_{3}^{}} \bigg) V_{N}^{\star}(\x(t)) + \eta \;,
	\label{eqn:Lyapunov_difference}
\end{align}
which implies that
\begin{align}
	\parallel \x(t+1) \parallel_{2}^{2} \leq \gamma\frac{a_{3}^{}}{a_{2}^{}} \parallel \x(t) \parallel_{2}^{2} + \frac{\eta}{a_{2}^{}} \;.
		\label{eqn:State_difference}
\end{align}
Since $0 < a_{2}^{} \leq a_{3}^{}$, it follows that $\gamma = 1 - \frac{a_{2}^{}}{a_{3}^{}}$, i.e., $0 < \gamma \leq 1$.
Therefore, by iterating~\eqref{eqn:State_difference}, it is possible to exponentially bound the state evolution via:
\begin{align*}
	\parallel \x(t) \parallel_{2}^{2} \leq \gamma_{}^{t} \frac{a_{3}^{}}{a_{2}^{}} \parallel \x(0) \parallel_{2}^{2} + \frac{1 - \gamma^{t}}{1 - \gamma}\frac{\eta}{a_{2}^{}} \;.
\end{align*}
Using the norm inequality $\parallel \v \parallel_{\infty}^{} \leq \parallel \v \parallel_{2}^{} \leq \sqrt{n} \parallel \v \parallel_{\infty}^{}$ for any $\v \in \mathbb{R}_{}^{n}$ (see~\cite{Ber:09}), we get:
\begin{align*}
	\parallel \x(t) \parallel_{\infty}^{2} \leq \gamma_{}^{t} n \frac{a_{3}^{}}{a_{2}^{}} \parallel \x(0) \parallel_{\infty}^{2} + \frac{1 - \gamma^{t}}{1 - \gamma}\frac{\eta}{a_{2}^{}} \;.
\end{align*}
Thus, $\displaystyle{\lim_{t\rightarrow\infty} \parallel \x(t) \parallel_{\infty}^{} \leq \mu}$. $\hfill\blacksquare$
%-------------------------------------------------------------------------------------------------------------------------------------------------------------------------------------------------------------------------
% Bibliografia
%-------------------------------------------------------------------------------------------------------------------------------------------------------------------------------------------------------------------------

\bibliographystyle{IEEEtran}
\bibliography{acc2013}

\end{document}